\documentclass[letterpaper,12pt]{article}

\usepackage{amsmath,amssymb,amscd,latexsym,dsfont,wasysym,bbm}
\usepackage{float}
\usepackage{color}
\usepackage{graphicx,subfigure}
\usepackage{multirow,multicol} 
\usepackage{verbatim}
\usepackage{psfrag}
\usepackage{comment}
\usepackage{makeidx}
\usepackage[]{algorithm}
\usepackage{algorithmicx}
\usepackage{algpseudocode}
\usepackage{cases}
 \usepackage{setspace}
\usepackage{pgfplots} 
\newcommand{\tr}[1]{\textrm{#1}}
\newcommand{\mr}[1]{\mathrm{#1}}

\newcommand{\mb}[1]{{\mathop{\boldsymbol{#1}}}}

\newcommand{\mc}[1]{\mathcal{#1}}
\newcommand{\mf}[1]{\mathsf{#1}}
 
\newcommand{\ms}[1]{\mathds{#1}}

\newcommand{\ov}[1]{\overline{#1}}

\newcommand{\balpha}{\boldsymbol{\alpha}}

\newcommand{\bdelta}{\boldsymbol{\delta}}

\newcommand{\btheta}{\boldsymbol{\theta}}

\newcommand{\figref}[1]{Fig.~\ref{#1}}

\newcommand{\secref}[1]{Sec.~\ref{#1}}

\newcommand{\tabref}[1]{Table~\ref{#1}}

\newcommand{\ie}{i.e.,~} 		
\newcommand{\eg}{e.g.,~}	

\newcommand{\argmax}{\mathop{\mr{argmax}}}

\newcommand{\set}[1]{\{#1\}}
\newcommand{\SET}[1]{\left\{#1\right\}}

\newcommand{\cd}{\cdot}
\newcommand{\ld}{\ldots}
\newcommand{\vd}{\vdots}

\newcommand{\e}{\mr{e}}

\newcommand{\PR}[1]{\Pr\SET{#1}}       	

\newcommand{\IND}[1]{\ms{I}\big[{#1}\big]}   	
\newcommand{\Ex}{\ms{E}}     			
\newcommand{\dd}{\,\mr{d}}             		

\newcommand{\mcS}{\mc{S}}

\newcommand{\mcY}{\mc{Y}}

\newcommand{\mfA}{\mf{A}}
\newcommand{\mfD}{\mf{D}}

\newcommand{\mfH}{\mf{H}}

\newcommand{\mfsH}{\mf{sH}} 
\newcommand{\mfwH}{\mf{wH}} 
\newcommand{\mfsA}{\mf{sA}} 
\newcommand{\mfwA}{\mf{wA}} 

\usepackage{dgjournal}

\usepackage[authoryear,comma,longnamesfirst,sectionbib,round]{natbib} 

\pgfplotsset{compat=1.12}
\usetikzlibrary{positioning,shadows,shapes,fit,arrows,backgrounds}

\tikzstyle{rect_my} = [draw, rectangle, minimum width=2cm, text width=1.8cm, fill=gray!15, 
  text centered,  minimum height=.9cm]
\tikzstyle{square_my} = [draw, rectangle, minimum width=1cm, text width=0.8cm, fill=gray!15, 
  text centered,  minimum height=.9cm]
\tikzstyle{square_my_graph} = [draw, rectangle, minimum width=1.2cm, text width=1cm, fill=gray!15, 
  text centered,  minimum height=1.2cm]
\tikzstyle{circle_my} = [draw, circle, minimum width=1cm, text width=0.8cm, fill=gray!15, 
  text centered,  minimum height=.9cm]
\tikzstyle{circle_my_graph} = [draw, circle, minimum width=1.1cm, text width=.8cm, fill=gray!15, 
  text centered]
\tikzstyle{cloud_my} = [draw, shape=cloud, minimum width=1cm, text width=0.8cm, fill=gray!15, 
  text centered,  minimum height=1cm]

\tikzstyle{point_my} = [draw=none, minimum width=0cm, text width=0cm, fill=none, 
  text centered,  minimum height=0cm]    
\tikzstyle{line_my} = [draw, -latex]    
\tikzstyle{box_my}=[draw, minimum size=2em, text width=4.5em, text centered]
\tikzstyle{bigbox_my}=[draw, inner sep=15pt]
\tikzstyle{arrow_my} = [thick,->,>=stealth]
\tikzstyle{noarrow_my} = [thick,-,=>stealth]

\usepackage[acronym,nonumberlist]{glossaries} 
\usepackage{url}

\newacronym[\glsshortpluralkey=PDFs,\glslongpluralkey=probability density functions]{pdf}{PDF}{probability density function}
\newacronym[\glsshortpluralkey=CDFs,\glslongpluralkey=cumulative density functions]{cdf}{CDF}{cumulative density function}
\newacronym[\glsshortpluralkey=CCDFs,\glslongpluralkey=complementary cumulative density functions]{ccdf}{CDF}{complementary cumulative density function}
\newacronym[\glsshortpluralkey=PMFs,\glslongpluralkey=probability mass functions]{pmf}{PMF}{probability mass function}
\newacronym[]{lhs}{l.h.s.}{left-hand side}
\newacronym[]{rhs}{r.h.s.}{right-hand side} 

\newacronym[]{bicm}{BICM}{bit-interleaved coded modulation}
\newacronym[]{bicmid}{BICM-ID}{BICM with iterative demapping}
\newacronym[]{cm}{CM}{coded modulation}
\newacronym[]{tcm}{TCM}{trellis-coded modulation}
\newacronym[]{mlc}{MLC}{multi-level coding}
\newacronym[]{pam}{PAM}{pulse amplitude modulation}
\newacronym[]{bpsk}{BPSK}{binary phase shift keying}
\newacronym[]{qam}{QAM}{quadrature amplitude modulation}
\newacronym[]{16qam}{16-QAM}{16-points quadrature amplitude modulation}
\newacronym[]{psk}{PSK}{phase shift keying}
\newacronym[\glsshortpluralkey=LLRs,\glslongpluralkey=logarithmic likelihood ratios]{llr}{LLR}{logarithmic likelihood ratio}
\newacronym[]{oc}{OC}{operating characteristic}
\newacronym[]{dmp}{DMP}{discretized message passing}
\newacronym[]{mp}{MP}{message passing}
\newacronym[]{ep}{EP}{expectation propagation}
\newacronym[\glsshortpluralkey=MIs,\glslongpluralkey=mutual informations]{mi}{MI}{mutual information}
\newacronym[\glsshortpluralkey=GMIs,\glslongpluralkey=generalized mutual informations]{gmi}{GMI}{generalized mutual information}
\newacronym[]{eesm}{EESM}{exponential effective-SNR-mapping}
\newacronym[]{bicm-gmi}{BICM-GMI}{BICM generalized mutual information}
\newacronym[]{awgn}{AWGN}{additive white Gaussian noise}
\newacronym[]{bsc}{BSC}{binary symetric channel}
\newacronym[]{amc}{AMC}{adaptive modulation and coding}
\newacronym[]{csi}{CSI}{channel state information}
\newacronym[]{cqi}{CQI}{channel quality indicator}
\newacronym[]{kl}{KL}{Kullback-Leibler}
\newacronym[]{cmm}{CMM}{circular moment matching}
\newacronym[]{ga}{GA}{Gaussian approximation}

\newacronym[]{sp}{SP}{set-partitioning}
\newacronym[]{gsm}{GSM}{global system for mobile communications}
\newacronym[]{edge}{EDGE}{enhanced data rates for GSM evolution}
\newacronym[]{3gpp}{3GPP}{3rd generation partnership project}
\newacronym[]{umts}{UMTS}{Universal Mobile Telecommunication System}
\newacronym[]{lte}{LTE}{Long Term Evolution}
\newacronym[]{dvb}{DVB}{digital video broadcasting}
\newacronym[]{fdd}{FDD}{Frequency Division Duplexing}

\newacronym[\glsshortpluralkey=CCs,\glslongpluralkey=convolutional codes]{cc}{CC}{convolutional code}
\newacronym[\glsshortpluralkey=PCCCs,\glslongpluralkey=parallel concatenated convolutional codes]{pccc}{PCCC}{parallel concatenated convolutional code}
\newacronym[\glsshortpluralkey=TCs,\glslongpluralkey=turbo codes]{tc}{TC}{turbo code}
\newacronym{ldpc}{LDPC}{low-density parity-check}
\newacronym[]{ofdm}{OFDM}{orthogonal frequency-division multiplexing}
\newacronym[]{bep}{BEP}{bit-error probability}
\newacronym[]{wep}{WEP}{word-error probability}
\newacronym[]{sep}{SEP}{symbol-error probability}
\newacronym[]{pep}{PEP}{pairwise-error probability}
\newacronym[]{ttcm}{TTCM}{turbo-trellis coded modulation}
\newacronym[]{uep}{UEP}{unequal error protection}
\newacronym[\glsshortpluralkey=CENCs,\glslongpluralkey=convolutional encoders]{cenc}{CENC}{convolutional encoder}
\newacronym[]{mimo}{MIMO}{multiple-input multiple-output}
\newacronym[\glsshortpluralkey=SNRs,\glslongpluralkey=signal-to-noise ratios]{snr}{SNR}{signal-to-noise ratio}
\newacronym[\glsshortpluralkey=SINRs,\glslongpluralkey=the signal-to-interference-plus-noise ratios]{sinr}{SINR}{the signal-to-interference-plus-noise ratio}
\newacronym[]{msb}{MSB}{most-significative bit}
\newacronym[]{bcjr}{BCJR}{Bahl--Cocke--Jelinek--Raviv}
\newacronym[]{cbc}{CBC}{Colavolpe--Barbieri--Caire}
\newacronym[]{skr}{SKR}{Shayovitz--Kreimer--Raphaeli}
\newacronym[\glsshortpluralkey=SEDs,\glslongpluralkey=squared Euclidean distances]{sed}{SED}{squared Euclidean distance}
\newacronym[\glsshortpluralkey=EDs,\glslongpluralkey=Euclidean distances]{ed}{ED}{Euclidean distance}
\newacronym[\glsshortpluralkey=MEDs,\glslongpluralkey=minimum Euclidean distances]{med}{MED}{minimum Euclidean distance}
\newacronym[]{core}{CoRe}{constellation rearrangement}
\newacronym[]{msd}{MSD}{multistage decoding}
\newacronym[]{pdl}{PDL}{parallel decoding of the individual levels}
\newacronym[\glsshortpluralkey=GCs,\glslongpluralkey=Gray codes]{gc}{GC}{Gray code}
\newacronym[]{brgc}{BRGC}{binary-reflected Gray code}
\newacronym[]{nbc}{NBC}{natural binary code}
\newacronym[]{fbc}{FBC}{folded-binary code}
\newacronym[]{bsgc}{BSGC}{binary semi-Gray code}
\newacronym[]{msp}{MSP}{modified set-partitioning}
\newacronym[]{ssp}{SSP}{semi set-partitioning}
\newacronym[]{fhd}{FHD}{free Hamming distance}
\newacronym[]{mfhd}{MFHD}{maximum free Hamming distance}
\newacronym[]{ods}{ODS}{optimal distance spectrum}
\newacronym[]{iud}{i.u.d.}{independent and uniformly distributed}
\newacronym[]{ud}{u.d.}{uniformly distributed}
\newacronym[]{iid}{i.i.d.}{independent, identically distributed}
\newacronym[]{ami}{AMI}{accumulated mutual information}
\newacronym[]{bico}{BICO}{binary-input continuous-output}
\newacronym[]{gh}{GH}{Gauss--Hermite}
\newacronym[]{gum}{GUM}{Gaussian--uniform mixture}

\newacronym[\glsshortpluralkey=BSs,\glslongpluralkey=base-stations]{bs}{BS}{base-station}
\newacronym[\glsshortpluralkey=MSs,\glslongpluralkey=mobile-stations]{ms}{MS}{mobile-stations}

\newacronym[]{phy}{PHY}{physical layer} 
\newacronym[]{rlc}{RLC}{Radio-Link control} 
\newacronym[]{ran}{RAN}{Radio Access Network} 
\newacronym[]{llc}{LLC}{logical link control} 
\newacronym[]{tcp}{TCP}{transmission control protocol} 
\newacronym[]{mac}{MAC}{media access control} 
\newacronym[]{fft}{FFT}{fast Fourier transform} 
\newacronym[]{ft}{FT}{Fourrier transform}
\newacronym[]{cf}{CF}{characteristic function} 
\newacronym[]{mgf}{MGF}{moment generating function} 
\newacronym[]{ee}{EE}{energy efficiency} 
\newacronym[]{eb}{EB}{energy per bit}
\newacronym[]{kkt}{KKT}{Karush--Kuhn--Tucker} 
\newacronym[]{mcs}{MCS}{modulation/coding scheme} 
\newacronym[]{fec}{FEC}{forward error correction}
\newacronym[]{arq}{ARQ}{automatic repeat request}
\newacronym[]{harq}{HARQ}{hybrid ARQ}
\newacronym[]{tarq}{TARQ}{truncated HARQ}
\newacronym[]{ir}{IR}{incremental redundancy}
\newacronym[]{rpr}{RR}{repetition redundancy}
\newacronym[]{rrharq}{RR-HARQ}{repetition redundancy HARQ}
\newacronym[]{irharq}{IR-HARQ}{incremental redundancy HARQ}
\newacronym[]{ack}{ACK}{positive acknowledgment}
\newacronym[]{nack}{NACK}{negative acknowledgment}
\newacronym[]{hol}{HoL}{head of the line}
\newacronym[]{crc}{CRC}{cyclic redundancy check}
\newacronym[]{dp}{DP}{dynamic programming}
\newacronym[]{gp}{GP}{geometric programming}
\newacronym[]{per}{PER}{packet error rate}
\newacronym[]{ber}{BER}{bit error rate}
\newacronym[]{op}{OP}{outage probability}
\newacronym[]{spa}{SPA}{saddle-point approximation}
\newacronym[]{mrc}{MRC}{maximum ratio combining}
\newacronym[]{mdp}{MDP}{Markov decision process}
\newacronym[]{lp}{LP}{linear programming}
\newacronym[]{pomdp}{POMDP}{partially observable Markov decision process}
\newacronym[]{psimdp}{PSI-MDP}{partial state information Markov decision process}
\newacronym[]{scpp}{SCPP}{stochastic shortest path problem}

\newacronym[]{forw}{frwd}{forward}
\newacronym[]{feed}{fdbk}{feedback}

\newacronym[]{mm}{MM-HARQ}{multi-message HARQ}
\newacronym[]{xp}{XP-HARQ}{cross-packet HARQ}
\newacronym[]{ts}{TS}{time-sharing}
\newacronym[]{sc}{SC}{superposition coding}
\newacronym[]{sbrq}{SBRQ}{systematic backward retransmission}
\newacronym[]{brq}{BRQ}{backward retransmission}
\newacronym[]{lharq}{L-HARQ}{layer-coded HARQ}
\newacronym[]{anlharq}{AoN-HARQ}{all-or-none L-HARQ}
\newacronym[]{vlharq}{VL-HARQ}{variable-length HARQ}

\newacronym[]{pp}{PPP}{point process}
\newacronym[]{ppp}{PPP}{Poisson point process}

\newacronym[]{fide}{FIDE}{F\'ed\'eration Internationale des \'Echecs}
\newacronym[]{fifa}{FIFA}{F\'ed\'eration Internationale de Football Association}
\newacronym[]{fivb}{FIVB}{F\'ed\'eration Internationale de Volleyball}
\newacronym[]{epl}{EPL}{English Premier League}
\newacronym[]{nhl}{NHL}{National Hockey League}
\newacronym[]{nfl}{NFL}{National Football League}

\newacronym[]{sg}{SG}{stochastic gradient}
\newacronym[]{lms}{LMS}{least mean squares}
\newacronym[]{rls}{RLS}{recursive least squares}
\newacronym[]{vss}{VSS}{variable step-size}
\newacronym[]{hfa}{HFA}{home-field advantage}
\newacronym[]{ha}{HA}{home advantage}
\newacronym[]{mov}{MOV}{margin of victory}
\newacronym[]{ac}{AC}{Adjacent Categories}
\newacronym[]{cl}{CL}{Cumulative Link}
\newacronym[]{rps}{RPS}{Ranked Probability Score}
\newacronym[]{mse}{MSE}{Mean Square Error}
\newacronym[]{mmse}{MMSE}{Minimum Mean Square Error}
\newacronym[]{rmse}{RMSE}{Root Mean Squares Error}
\newacronym[]{map}{MAP}{maximum a posteriori}
\newacronym[]{ml}{ML}{maximum likelihood}
\newacronym[]{loo}{LOO}{leave-one-out}
\newacronym[]{alo}{ALO}{approximate leave-one-out}

\newacronym[]{svd}{SVD}{singular values decomposition}

\newacronym[]{skf}{SKF}{Simplified Kalman Filter}
\newacronym[]{vskf}{vSKF}{\emph{vector-covariance} Simplified Kalman Filter}
\newacronym[]{sskf}{sSKF}{\emph{scalar-covariance} Simplified Kalman Filter}
\newacronym[]{fskf}{fSKF}{\emph{fixed-variance} Simplified Kalman Filter}
\newacronym[]{kf}{KF}{Kalman Filter}
\newacronym[]{gelo}{G-Elo}{Generalized Elo}

\newacronym[]{tpb}{TPB}{tensor-product-basis}

\begin{document}

\title{G-Elo: Generalization of the Elo algorithm \\ by modeling the discretized Margin of Victory}
\author{Leszek Szczecinski
\thanks{%
L.~Szczecinski  is with Institut National de la Recherche Scientifique, Montreal, Canada. [e-mail: Leszek.Szczecinski@inrs.ca].}%
}%


\originalmaketitle  

\setstretch{1.6} 

\begin{abstract} 
In this work we develop a new algorithm for rating of teams (or players) in one-on-one games by exploiting the observed difference of the game-points (such as goals), also known as a \gls{mov}. Our objective is to obtain the Elo-style algorithm whose operation is simple to implement and to understand intuitively.  This is done in three steps: first, we define the probabilistic model between the teams' skills and the discretized \gls{mov} variable: this generalizes the model underpinning the Elo algorithm, where the \gls{mov} variable is discretized into three categories (win/loss/draw). Second, with the formal probabilistic model at hand, the optimization required by the maximum likelihood rule is implemented via stochastic gradient; this yields simple on-line equations for the rating updates which are identical in their general form to those characteristic of the Elo algorithm: the main difference lies in the way the scores and the expected scores are defined. Third, we propose a simple method to estimate the coefficients of the model, and thus define the operation of the algorithm; it is done in a closed form using the historical data so the algorithm is tailored to the sport of interest and the coefficients defining its operation are determined in entirely transparent manner. The alternative, optimization-based strategy to find the coefficients is also presented. We show numerical examples based on the results of the association football of the English Premier League and the American football of the National Football League.
\end{abstract}

\section{Introduction}\label{Sec:Intro}
This work is concerned with the rating of teams in one-on-one games taking into account the \acrfull{mov} which may be defined, for example, by the difference of the scored game-points (such as goals). Our objective is to obtain a simple, Elo-style algorithm which should be based on a formal probabilistic model between the \gls{mov} variable and the skills/strengths of the teams. 

Rating of the teams is one of the fundamental problems in sport analytics and consists in finding a numerical value which reflects each team's skill/strength (the same approach is also applicable to individual sports, in which case we talk about \emph{players} rather than teams). The conventional approach, used in many sports relies on giving each team a fixed number of league-points for a particular game outcome (\eg in association football, three league-points are given for a win, one point for a draw, and zero for a loss). The league-points are defined in a sport-specific manner, supposedly reflecting the difficulty of reaching a particular game outcome.

More advanced rating strategies rely on probabilistic models which link the skills to the outcomes. The rating consists then in inferring the teams' skills (\ie the unobservable parameters) from the observed outcomes of the games. There are then two modeling questions that must be answered i) how to model the relationship between the outcome of the game and the skills, and ii) how to model the dynamics of the skills, \ie the relationship between the skills across games. Each of these modeling issues requires finding the coefficients which define the respective models. In this work we are mostly concerned with the first question while the dynamics is considered indirectly via stochastic optimization; more on this issue, later.

Many models and resulting algorithms have been proposed in the literature and, while very few have been adopted in practice, undoubtfully the most successful rating algorithm was devised by Arpad Elo to rate chess players \citep{Elo78_Book}. Adopted by the \gls{fide} in the sixties and, recently, by the \gls{fifa} for the rating of the national football teams \citep{fifa_rating}, it was also used in other sports, although mostly informally, \eg in American football by \citet{Silver20}, as well as in eSports \citep{Herbrich06}. 

The Elo algorithm can be written as 
\begin{align}\label{Elo.basic}
   \theta \leftarrow \theta +  K \big( y   - G(z)\big),
\end{align}
where $\theta$ is the skill of the player and it is updated after the game (hence the symbol ``$\leftarrow$''); the amplitude of the update depends on a step $K$, the score of the game, $y\in\set{0,\frac{1}{2},1}$ (corresponding to the loss, the draw and the win) and on the expected score $G(z)\in(0,1)$ which is calculated using $z=\theta-\theta'$, where $\theta'$ is the skill of the opposing player.

It can be argued that the popularity of the Elo algorithm was earned due to (i) the simplicity: the rating rules
\eqref{Elo.basic} can be easily implemented), (ii) the interpretability: the amplitude of the rating update is proportional to the difference between the observed and the expected scores of the game (expressed by $y - G(z)$ in \eqref{Elo.basic}), and (iii) a sport independent operation: the algorithm has no coefficients which take the sport-specific information into account;
thus, it can – and has been applied in different competitions with ternary, win/draw/loss game results.

The simplicity of the Elo algorithms is indisputably appealing but the algorithm is sport-independent only in the binary games, where there is no coefficients to adjust (and even then, we have to deal with the \gls{hfa} in a sport-specific manner). On the other hand, the sport-independence is merely apparent in the ternary games: as shown in \citet{Szczecinski20}, the Elo algorithm is implicitly based on the draw model proposed by \citet{Davidson70} and assumes a particular value of the draw coefficient that can be related to frequency of draws equal to 50\%.  This is clearly unrealistic in most sports so, to correct for this unrecognized algorithmic bias, \citet{Szczecinski20} proposed the \emph{Elo-Davidson algorithm} which is as simple and interpretable as the Elo algorithm, yet  it provides the possibility to set the model's coefficients taking into account both, the frequency of draws as well as the \gls{hfa}, which are sport-dependent.

\subsection{Margin of victory in the rating algorithms}\label{Sec:MOV.literature}
The Elo and the Elo-Davidson algorithms, however, are defined for ternary games and thus are not suited for operation in the presence of the multilevel \gls{mov} variable/outcome, $d$, defined by the difference in game-points. This is the problem we want to address here. 

To put it into perspective, we briefly outline the venues which were previously taken to deal with the \gls{mov}:
\begin{enumerate}
\item {\bf Direct modeling}, where $d$ is assumed to be drawn from a predefined distribution. For example, the Gaussian distribution is used in \citet{Massey97},\footnote{This is done implicitly though: \citet{Massey97} uses least-squares criterion to fit the rating to the observations and it is easy to see that this corresponds to the assumption of the Gaussian distribution of the \gls{mov} value, $d$.}  while \citet{Karlis08} and \citet{Manderson18} use the Skellam distribution.

The difficulty is that, there is no guarantee that any particular sport yields $d$ with the distribution one deems suitable for the rating and, usually, there is not much room to adjust the model. This issue becomes immediately apparent when looking at the number of coefficients defining the distribution. In fact, there is none if we use the Gaussian distribution (because the value of the variance is irrelevant to the fit) and one coefficient may be adjusted in the Skellam distribution. 

This lack of flexibility may result in serious discrepancies between the model and the observations. For example, to deal with the high frequency of draws in association football (that is, events when $d=0$), a hybrid model which is a mixture of a the Skellam distribution and the mass probability function related to the frequency of draws was introduced in \citet[Sec.~2.3]{Karlis08}. However, such hybrid models lead to optimization problems that may be not convex and the resulting algorithms are difficult to apply to on-line results.\footnote{For example, \citet{Karlis08} uses Markov chain Monte Carlo method to optimize the parameters which is quite complex when compared to a very simple and well-known Elo algorithm \citep{Elo78_Book}.}

\item {\bf Discretization} consists in transforming $d$ into categorical variables, for which we have to find a suitable model. Since the categorical data modeling usually relies on a large number of coefficients, this approach provides more flexibility than the direct modeling of $d$. Using the so-called  \gls{cl} model \citep[Ch.~8.3]{Agresti13_book}, and which generalizes the model proposed for ternary games by \citet{Rao67}, this approach was applied, \eg  in \citet{Fahrmeir94}, \citet{Knorr00}, \citet{Held05}, \citet{Goddard05}.

The principal difficulty is that the coefficients of the model must be estimated from the data. However, in our view, the main issue is that the resulting algorithms do not yield a simple update as the one we show in \eqref{Elo.basic} and thus are more difficult to understand/interpret.

\item {\bf Fine-tuning of the Elo algorithm} takes advantage of the adopted Elo algorithm defined in \eqref{Elo.basic} whose operation is modified using observed \gls{mov} variable, $d$.  For example, in \citet{Hvattum10}, \citet{Silver20}, or in \citet{Kovalchik20}, the amplitude  of the rating update---the variable $K$ in \eqref{Elo.basic}---is made dependent on  $d$.

Since these works did not rely on the explicit probabilistic model relating the skills and the multilevel outcomes $d$,  they were treated as heuristics \citep[Sec.~3.1.4]{Kiraly17}. With that regard, the recent work by \cite{Ingram21} is different and has shown a formal model linking the amplitude $K$ with the \gls{mov} variable $d$. It requires the Bayesian assumption, that is, the skills are not the points estimates but are assumed to be normally distributed with a (fixed) posterior variance.

\item {\bf Multidimensional skills} are used to model different aspects of the game and may be related to the game-points scored by both teams (and not only their difference, $d$). In particular,  the offensive and defensive skills may be estimated as done for example in \citet{Maher82}, \citet{Dixon97}, or in \citet{Boshnakov17}. 

However, since the skills are then two-dimensional, they cannot be directly compared and thus, it is not obvious whether it is possible to  define the notion of rating as we understand it here, and which should imply a possibility of ordering of the skills. So, while the multi-dimensional modeling of skills is interesting in its own right, it is also a different issue and we will not deal with it in our work.
\end{enumerate}

\subsection{Contributions}\label{Sec:Contributions}
The presented overview indicates that the algorithm which inherits the simplicity of the Elo algorithm shown in \eqref{Elo.basic}, and that considers the \gls{mov} in a formal probabilistic framework, has not yet been proposed with a notable exception of the recent derivations shown in  \citet{Ingram21}.

To devise such an algorithm we have to find a  model which relates the skills and the multi-level outcomes represented by the \gls{mov} variable, $d$. Our take on this issue is different from the one presented in \citet{Ingram21}, where $d$ is modeled as a Gaussian variable. In our view, the most natural way of generalizing the Elo rating algorithm, is by recognizing first that the Elo (and the Elo-Davidson) algorithm is based on a particular case of the \gls{ac} model which is well-known in the literature on the ordinal data analysis \citep[Ch.~8.3]{Agresti13_book}, \citep[Sec.~3]{Agresti92}. In the ternary-output case, the \gls{ac} model corresponds to the one proposed in \citet{Davidson70}--a relationship which was recognized in the literature, see  \citet{Sinclair82}, \citet{Dittrich07}, and \citet{Szczecinski20}, but---to the best of our knowledge---was not exploited in the sport-rating literature for games with higher-than-ternary outcomes.

The \gls{ac} model relies on the discretization of the \gls{mov} variable, $d$, into \emph{categories} which are ordinal variables. This is a general idea already  applied before but using   the \gls{cl} model rather than the \gls{ac} model we want to use. This distinction is important because, even if the differences between the predictive power of the \gls{cl} and \gls{ac} models do not seem to be critical, see \citet[Ch.~8.3.4]{Agresti13_book}, by adopting  the \gls{ac} model we will obtain very simple,  Elo-style equations for rating updates; it is not the case when the \gls{cl} model is used, see \citet[Corollary~1]{Szczecinski20}. 

In fact,  the equations of the \gls{gelo} rating algorithm we propose in this work are the same as those defining the Elo algorithm in \eqref{Elo.basic}, except that the score and the expected score (corresponding to the terms, $y$ and $G(z)$ in \eqref{Elo.basic}) are redefined. This is our most important contribution. We note again that our approach is model-based and thus, the changes in the score and the expected score are due to our choice of the \gls{ac} model. This is worth emphasizing because it is possible to change the score definition without defining a new model; the \gls{fifa} rating is the best example of such a heuristic.

The second contribution appears in the way we find the coefficients of the \gls{ac} model: they are obtained in a closed-form from the frequencies with which each category of $d$ is observed in the historical data. This follows from the ideas proposed in \citet{Szczecinski20}, and allows us to define the model and the algorithm in a simple, sport-specific, and entirely transparent manner. We discuss this strategy and compare it to the more conventional, optimization-based methods.

The modeling approach we adopt is focused on the skills-outcomes relationship and to estimate the skills we apply the \gls{ml} principle. To avoid a rigidity of the assumption of skills being constant throughout the games, we use the \gls{sg} optimization that can be also related to the  Elo algorithm. It allows us to update the skills in real-time, as well as, to follow the changes in the teams' skills. So, while there is no explicit model defining the dynamics of the skills, this phenomenon is  taken into account implicitly. At the cost of increased complexity of the algorithm, it is possible to augment the model and define the skills' dynamics using a random walk, as done before, \eg in \citet{Fahrmeir94}, \citet{Glickman99}, \citet{Held05}, or in \citet{Manderson18}.

The paper is organized as follows. The problem is defined in \secref{Sec:Model} where we propose a new model and derive the algorithm. In \secref{Sec:Coeffs} we discuss how the coefficients of the model can be estimated, while \secref{Sec:Examples} provides numerical examples obtained for the data taken from the \gls{epl} and the \gls{nfl} seasons. The conclusions are drawn in \secref{Sec:Conclusions}.

\section{Modeling and rating in multi-level games}\label{Sec:Model}
We consider a scenario where $M$ teams, indexed by $m=1,\ld,M$, challenge each other in face-to-face games. The games are indexed with $t$ and involve the teams defined by the pair of indices $(i_{t},j_{t})$. The index $i_{t}$ refers to the ``home'' team, while $j_{t}$ indicates  the ``away'' team. This distinction allows us to take into account the \acrfull{hfa}; we address this issue later. The result/outcome of the game is denoted by $y_t$; in practice, many games, \eg indexed by $t, t+1, t+2$, may be played simultaneously but we are not concerned with this issue as the temporal relationships are not explicitly modeled here. 

Having observed the outcomes of the games, $y_l, l=1,\ld,t$, we want to assign real numbers known as \emph{ratings} or \emph{skills}, \citep{Herbrich06}, \citep{Caron12} to each of the teams involved in the games. The skills are gathered in the vector $\btheta_t=[\theta_{t,1},\ld,\theta_{t,M}]$, where the indexing with $t$ is necessary because the skills may evolve in time. Thus, at the game $t$, the skill of the teams facing each other are $\theta_{t,i_t}$ (for the home team) and $\theta_{t,j_t}$ (for the away team); knowing $\theta_{t,i_t}$ and $\theta_{t,j_t}$ we should be able to infer the probability of obtaining a particular game outcome $y_t$ (\eg draw). After the outcome $y_t$ is observed, the re-estimated (updated) skills are placed in $\btheta_{t+1}$.

We assume that the results $y_t$ belong to an ordinal $(J+1)$-ary set. For example,  in many team sports, such as football, hockey, or basketball the outcomes are taken from a ternary set comprising  the result of ``home win'' ($\mfH$), ``away win'' ($\mfA$), or ``draw'' ($\mfD$); thus $J=2$. 

Defining the outcomes in such a way may be seen as a ternary discretization of the difference between the games-points we denote by $d_t$ (\eg the goal difference, in the team sports we mentioned). That is,
\begin{align}\label{3levels}
\set{d_t<0}&\equiv \mfA,\quad
\set{ d_t=0}\equiv \mfD,\quad
\set{0< d_t}\equiv \mfH.
\end{align}

However, considering only the sign of $d_t$ (and $d_t=0$) may hamper our ability to rate the teams; this corresponds to our intuition, that winning by a small margin, \eg one game-point, $d_t=1$, is not the same as winning by a large margin, \eg  $d_t=3$. The idea here is thus to use the valuable information contained in the \gls{mov} variable, $d_t$, and go beyond the ternary discretization defined in \eqref{3levels}.

In the probabilistic perspective, exploiting the \gls{mov}  amounts to defining the model between $d_{t}$, and the ratings of the involved teams, $\theta_{t,i_t}$ and $\theta_{t,j_t}$, and in our work we will propose a rating strategy that i) formally models the multiple outcomes of the game using a probabilistic approach, and ii) ensures the transparency and the simplicity of the rating, generalizing in fact the well-known Elo rating algorithm.

\subsection{Discretization and adjacent categories model}\label{Sec:AC.model}
We start addressing the issue of the multilevel results by discretizing the \gls{mov} variable, $d_t$. As a running example of the multi-level outcome, and to go beyond the ternary discretization of the difference between scored points, we define the results as follows:
\begin{align}
\label{res.sA}
\set{d_t<-\Delta}&\equiv \mfsA \quad\text{(strong away win)} &&\implies&&  y_t=0\\
\label{res.wA}
\set{-\Delta\leq d_t<0}&\equiv \mfwA\quad\text{(weak away win)}&&\implies&&  y_t=1\\
\label{res.D}
\set{ d_t=0}&\equiv\mfD\quad \text{(draw)}&&\implies&&  y_t=2\\
\label{res.wJ}
\set{0< d_t\leq \Delta}&\equiv \mfwH\quad\text{(weak home win)}&&\implies&&  y_t=3\\
\label{res.sJ}
\set{\Delta< d_t}&\equiv \mfsH\quad\text{(strong home win)}&&\implies&&  y_t=4.
\end{align}
In this case $J=4$ but there is no intrinsic limit on the value of $J$. The threshold $\Delta$ must yet be defined and it may be fixed via optimization or determined by the experts of the game. But, even if we do not know how it was decided, everyone will understand its meaning from the above definitions. In general, the threshold, $\Delta$, should depend on the sport/competition as there is no reason to believe that the point difference defining the weak win should be the same in all sports; this will becomes clear in the examples comparing the association- and the American football games. 

Comparing to the direct modeling of $d_t$ as a random variable, the strength of the above discretization of $d_t$ into $y_t$ is that we can focus on the most relevant results and will not be affected by the rare events, \ie the outliers, such as large values of $|d_t|$.

Furthermore, not only the discretization may be adjusted using the sport-specific knowledge but the discretization categories allow us to combine the objective measures, such as game-points (goals) with a subjective evaluation of the game.\footnote{For example, in ice hockey, in the last minute(s) of the game, the losing team often replaces the goalie with an offensive skater to increase the chances of drawing (which, in turn, forces the overtime). However, this also leaves the net undefended letting the winning team to score. While this potentially increases the \gls{mov}, $d_t$, it is not necessarily a sign of an overwhelming superiority of the winning team. Then, knowing that the last goal was scored into the empty net, we might reclassify the event, \eg from $\mfsH$ to $\mfwH$. 

Similarly, we might reconsider the classification of the results knowing that the goal was scored due to an obvious referee's judgement error, \eg declaring the penalty shot in football, which changes the \gls{mov} variable $d_t$ but which we may consider unjustified.}

The game results are thus ordinal in nature and, for convenience of notation we use numerical indicators assigned to the outcomes $y_t$ and defined over the set $\set{0,\ld,J}$. The encoding into integer indices is arbitrary (later we show how to change it) but is largely used in the literature and eases the operations. For example, it allows us to express the natural symmetry of the result: switching the position of the home/away teams would require changing of the result into \mbox{$y^\tr{switch}_t= J-y_t$}, \eg the strong home win, $y_t=J$,  would become a strong away win,  \mbox{$y^\tr{switch}_t=0$}.

The modeling of the ordinal variables is well studied in the literature, \eg in \citet[Ch.~8.2-8.3]{Agresti13_book} and we use here the \gls{ac} model \citep[Sec.~3]{Agresti92} which may be summarized as follows:
\begin{align}\label{Pr.y.t}
\PR{Y_t=h | z_t}\propto  10^{\alpha_h +\delta_h\frac{z_t}{\sigma}}, \quad h=0,\ld, J,
\end{align}
where $\alpha_h, \delta_h, h=0,\ld, J$ are the coefficients of the model, the skills affect the probability via their difference (a common assumption in the rating algorithms)
\begin{align}\label{z.t}
z_t = \theta_{t,i_t} - \theta_{t,j_t},
\end{align}
and $\sigma>0$ is an arbitrarily set \emph{scale}; it is often used in sport rating to ``stretch'' (the range of) the skills; for example, \gls{fifa} uses $\sigma=600$ and \gls{fide}, $\sigma=400$.\footnote{Changing of the scale may also be interpreted as a change of the base of the exponential in \eqref{Pr.y.t}; for example, we may write: $10^{\delta_h \frac{z_t}{\sigma}}=\e^{\delta_h \frac{z_t}{\sigma'}}$ with $\sigma'=\sigma\log_{10}(\e)$. The inverse is also true: changing the base corresponds to changing the scale.}

A direct consequence of the model defined by  \eqref{Pr.y.t} is that 
\begin{align}
\log_{10} \frac{ \PR{Y_t=h|z_t} }{ \PR{Y_t=l|z_t} } = (\alpha_{h}-\alpha_{l}) + (\delta_h-\delta_l) \frac{z_t}{\sigma},
\end{align} 
so the \gls{ac} model ensures that the differences between the log-probabilities of the results grow linearly with the difference of the rating levels, $z_t$, provided $\delta_h$ grows monotonically with $h$:\footnote{The requirement $\delta_h>\delta_{h-1}$, may be seen as a formal condition for treating the variables $y_t$ as ordinal (as opposed to nominal). } for example, with growing difference $z_t$, the probability of  a ``strong home win''  approaches one, \ie $\lim_{z_t\rightarrow\infty}\PR{\mfsH }=1$.

The normalization of \eqref{Pr.y.t} yields
\begin{align}\label{Pr.y.t.normalized}
\PR{Y_t=h|z_t}
=
P_h(z_t)  
=
\frac{  10^{\alpha_h+\delta_h\frac{z_t}{\sigma}}}
{\sum_{l=0}^J  10^{\alpha_l + \delta_l\frac{z_t}{\sigma}}}, \quad h=0,\ld, J,
\end{align}
which is a multinomial logistic model \citep[Sec.~8.1.3]{Agresti13_book}. 

A convenient symmetry  is obtained assuming momentarily that the results are independent of the teams' order allowing us to switch their indices,\footnote{In other words, in the absence of the \acrfull{hfa}.} \citep[Sec.~2.1]{Fahrmeir94}, \citep[Sec.~3]{Agresti92}, which yields  \mbox{$P_h(z_t)=P_{J-h}(-z_t)$}. Since this should hold for all $z_t$, we require that
\begin{align}\label{alpha.delta}
\alpha_h&=\alpha_{J-h},\qquad\delta_h=-\delta_{J-h}. 
\end{align}

Furthermore, replacing $\alpha_h$ by $\alpha+\alpha_h, h=0,\ld, J$ and/or $\delta_h$ by $\delta+\delta_h, h=0,\ld, J$ does not change \eqref{Pr.y.t.normalized}. Therefore, without loss of generality, we can fix one $\alpha_h$ and one $\delta_h$; here, somewhat arbitrarily, we decide to set $\alpha_0=\alpha_J=0$ and $\delta_J=-\delta_0=1$ so, to define \eqref{Pr.y.t.normalized}, $J-1$ independent parameters are required.\footnote{For $J$ even (\ie with odd number of categories), we need $\frac{J}{2}$ coefficients $\alpha_h, h=1,\ld, \frac{J}{2}$ and $\frac{J}{2}-1$ coefficients $\delta_h, h=1, \ld, \frac{J}{2}-1$, where $\delta_{\frac{J}{2}}=0$ due to \eqref{alpha.delta}. For $J$ odd, we must  define  $\alpha_h, \delta_h, h=0,\ld, \frac{J-1}{2}$.}

While the choice of the model is arbitrary,\footnote{As we said in \secref{Sec:Intro}, the \gls{cl} model, \citep[Sec.~2]{Agresti92}, is the most popular choice in the sport rating literature, \eg in \citet{Fahrmeir94} or in  \citet{Knorr00}.} the \gls{ac} model defined by \eqref{Pr.y.t.normalized} has the advantage of being a generalization of the Bradley-Terry (if $J=1$) and the Davidson (if $J=2$) models which are known to underpin the Elo rating algorithm \citep{Szczecinski20}. 

\subsection{Rating algorithm}\label{Sec:rating.filtering}

Rating consists in infering the parameters $\btheta_t$ from the game outcomes $y_{1},\ld, y_{t}$ using the probabilistic model $\PR{Y_t=y_t|z_t}$ we have shown in the previous section. 

Following \citet{Szczecinski20}, we derive the rating algorithm in two steps: first, we define the objective function to be optimized; this is the log-likelihood of the observation $y_t$ conditioned on the ratings $\btheta_t$. Next,  we apply the \acrfull{sg}  optimization which yields an approximate \acrfull{ml} solution to the estimation problem.

The log-likelihood of the result $y_t$ is calculated directly from \eqref{Pr.y.t.normalized}
\begin{align}
\label{likelihood}
L_t(\btheta_t)&=\log\PR{Y_t=y_t|\btheta_t}=\log\PR{Y_t=y_t|z_t}\\
\label{likelihood.2}
&=\log( 10)[\alpha_{y_t} + \delta_{y_t}\frac{z_t}{\sigma}] -\log \sum_{l=0}^J 10^{\alpha_l+\delta_{l}\frac{z_t}{\sigma}};
\end{align}
whose derivative with respect to $z_t$ is given by
\begin{align}
\label{dL.dz}
\frac{\dd}{\dd z_t}L_t(\btheta_t)
&= 
\frac{\log( 10)}{\sigma}
\left[
\delta_{y_t}
- \frac
{\sum_{l=0}^J  \delta_l \cd 10^{\alpha_l+\delta_l\frac{z_t}{\sigma}}}
{\sum_{l=0}^J 10^{\alpha_l+\delta_l\frac{z_t}{\sigma}}}
\right]\\
\label{dL.dz.1}
&= 
\frac{\log( 10)}{\sigma}
\left[
(\delta_{y_t}-\delta_0)
- \frac
{\sum_{l=0}^J  (\delta_l-\delta_0) \cd 10^{\alpha_l+\delta_l\frac{z_t}{\sigma}}}
{\sum_{l=0}^J 10^{\alpha_l+\delta_l\frac{z_t}{\sigma}}}
\right]\\
\label{dL.dz.2}
&= 
\frac{2\log( 10)}{\sigma}
\left[
\tilde y_t
- G(z_t)
\right],
\end{align}
where
\begin{align}
\label{y.t}
\tilde y_t &= \tilde \delta_{y_t},\\
\label{delta.tilde}
\tilde \delta_{h}&= \frac{1}{2}(\delta_{h}-\delta_0), \\
\label{G.z.c}
G(z)&=
\frac
{\sum_{l=0}^J \tilde \delta_l \cd 10^{\alpha_l+2\tilde{\delta}_l\frac{z}{\sigma}}}
{\sum_{l=0}^J  10^{\alpha_l+2\tilde{\delta}_l\frac{z}{\sigma}}}.
\end{align}

We note that we can treat $\tilde y_t$ as a new result of the game which takes values in the set $\tilde\mcY=\set{0, \tilde \delta_1, \ld, \tilde\delta_{J-1}, 1}$, where we used the obvious relationships $\tilde\delta_0=0$ and $\tilde\delta_J=1$ (implied by \eqref{delta.tilde} and by $\delta_h=-\delta_{J-h}$). The shifting and rescaling of the outcomes (\ie the transformation from the integers $y_t$ to fractions $\tilde{y}_t$ defined in \eqref{delta.tilde}) allows us to treat $\tilde{y}_t$ as a ``score'' of the game $\tilde{y}_t\in[0,1]$, exactly as it is the case in the Elo algorithm. 

From the antisymmetry $\delta_h=-\delta_{J-h}$ and from  Eq.~\eqref{delta.tilde} we obtain
\begin{align}\label{sym.tilde.delta.h}
\tilde\delta_{J-h}=1-\tilde\delta_{h},
\end{align}
which also means that only in the binary and ternary games the score set is predefined, respectively, as $\tilde\mcY=\set{0,1}$ and $\tilde\mcY=\set{0,\frac{1}{2},1}$. On the other hand, for $J>2$ we must first find (predefine or estimate) the coefficients $\delta_h, h=1,\ld,J-1$ and they, in turn, define the scores $\tilde\mcY=\set{0,\tilde\delta_1,\ld,\tilde\delta_{J-1},1}$. We  show how to do it in \secref{Sec:Coeffs}.


We are now ready to find the skills' update equations: from $z_t = \theta_{t,i_t} - \theta_{t,j_t}$ we get \mbox{$\frac{\dd L_t(\btheta_t)}{\dd \theta_{t,i_t}}=\frac{\dd L_t(\btheta_t)}{\dd z_t}$} and $\frac{\dd L_t(\btheta_t)}{\dd \theta_{t,j_t}}=-\frac{\dd L_t(\btheta_t)}{\dd z_t}$. The optimization via the \gls{sg} algorithm consists in changing the variables $\theta_{t,i_t}$ and $\theta_{t,j_t} $ in the direction indicated by the derivative which yields the following equations defining the \gls{gelo} algorithm:
\begin{align}
\label{algo.i}
\theta_{t+1,i_t} & \leftarrow \theta_{t,i_t} + \tilde{K} \sigma \big( \tilde y_t  -G(z_t) \big),\\
\label{algo.j}
\theta_{t+1,j_t} & \leftarrow \theta_{t,j_t} - \tilde{K} \sigma \big( \tilde y_t -G(z_t) \big),
\end{align}
where $\tilde{K}$ is the adaptation step which absorbs the constant terms from \eqref{dL.dz.2} and the multiplication by $\sigma$ makes the results independent of $\sigma$; that is, for a given $\tilde{K}$, the normalized skills $\theta_{t,i_t}/\sigma$ do not depend on $\sigma$.\footnote{We may thus set $\sigma=1$, run the algorithm, and multiply the results by $\sigma$ at the end.} As usual,  the teams which do not play in the game $t$ keep theirs skills unaltered,  \ie $\theta_{t+1,n}\leftarrow \theta_{t,n}$ for $n\notin\set{i_t,j_t}$.

At the beginning of each season, the skills may be initialized to a fixed value $\theta_{0,i}=\theta_{\tr{init}}$, but from \eqref{algo.i}-\eqref{algo.j} it is obvious that the initialization merely offsets the results because $z_t$ is insensitive to the common offset value, $\theta_{\tr{init}}$.

Further, denoting by $\tilde{Y}_t$ the random variable modeling the scores $\tilde{y}_t \in\tilde\mcY$, we see that 
\begin{align}\label{Ex.Y.tilde}
\Ex[ \tilde{Y}_t | z_t]  =  \sum_{\tilde y \in\tilde\mcY} \tilde{y} \PR{\tilde{Y}_t=\tilde{y}|z_t} = \sum_{h=0}^J \tilde{\delta}_h \PR{Y_t=h|z_t}= G(z_t )
\end{align}
so $G(z)$ has the meaning of the expected score. The fact that the difference between the score and the expected score is proportional to the  derivative of the log-likelihood, see \eqref{dL.dz.2}, is a consequence of the \gls{ac} model we adopted.

Also, noting that $G(-z)=1-G(z)$ it is easy to see that \eqref{algo.i}-\eqref{algo.j} may be concisely represented as
\begin{align}
\label{algo.ij}
\theta_{t+1,i} & \leftarrow \theta_{t,i} + \tilde{K}\sigma \big( \tilde y  -G(z) \big),
\end{align}
where $i\in\set{i_t,j_t}$ is the index of the team whose rating we want to change, and, denoting by
$j\in\set{i_t,j_t}$ the index of its opponent, \ie $j\neq i$, the difference in the rating levels of the teams before the game is denoted by $z=\theta_i-\theta_j$. The score $\tilde y$ is given by $\tilde y =\tilde y_t$ if $i=i_t$ (\ie we rate the home team), and $\tilde y =1- \tilde y_t$ if the team $i=j_t$ (we rate the away team). 


\subsection{Relationship to Elo/Elo-Davidson rating algorithms}\label{Sec:relationship}

The updates of the \gls{gelo} algorithm defined in \eqref{algo.ij} have exactly the same form as those of the conventional Elo algorithm (as well as, the Elo-Davidson algorithm) defined in \eqref{Elo.basic}. Thus, the \gls{gelo} algorithm can be considered as a true generalization of the Elo rating algorithm being able to deal with arbitrarily defined \gls{mov}. In  the case of binary and ternary games,  the Elo and the Elo-Davidson algorithms are special instances of the \gls{gelo} rating. In particular:
\begin{itemize}
\item 
For the binary games, \ie with $J=1$, Eq.~\eqref{G.z.c} is reduced to 
\begin{align}\label{G.z.Elo}
G(z)&=
\frac{ 10^{\frac{z}{\sigma}}}
{10^{-\frac{z}{\sigma}}+10^{\frac{z}{\sigma}}}
\end{align}
and we attribute the score $\tilde{y}=1$ to the winning, and $\tilde{y}=0$ to the losing team. Thus we recover the conventional Elo rating algorithm for the binary games. 
\item 
For ternary games, $J=2$ and we augment the score space $\tilde\mcY$ to include $\tilde{y}=\frac{1}{2}$ for the draw,  and Eq.~\eqref{G.z.c} becomes
\begin{align}\label{G.z.Davidson}
G(z)&=
\frac{ 10^{\frac{z}{\sigma}} +\frac{1}{2}\kappa}
{10^{-\frac{z}{\sigma}}+ 10^{\frac{z}{\sigma}}+\kappa},
\end{align}
which corresponds to the Elo-Davidson algorithm \citep[Sec.~3]{Szczecinski20} with $\kappa=10^{\alpha_1}$.

It is worthwhile to note that setting $\kappa=2$ we obtain
\begin{align}\label{G.z.Davidson.2}
G(z)&=
\frac{ 10^{\frac{z}{\sigma}} +1}
{10^{-\frac{z}{\sigma}}+ 10^{\frac{z}{\sigma}}+2}
=\frac{10^{\frac{z}{2\sigma}}}{10^{-\frac{z}{2\sigma}}+ 10^{\frac{z}{2\sigma}}},
\end{align}
which, after replacement $\sigma \leftarrow \sigma/2$ is the same as the expected score for the binary games  in \eqref{G.z.Elo}. In such a case, moving from the binary to the ternary games, the only change required in the algorithm is to use the fractional score for the draw (\ie $\tilde{y}_t=\frac{1}{2}$). This explains why the transition from the binary to ternary games is seemingly effortless in the Elo algorithm.

However, as mentioned in \secref{Sec:Intro} there is a cost associated with this apparent simplicity: not only we hide the ternary-outcomes model but also, using (implicitly) the value $\kappa=2$ corresponds approximately to the assumption of $50\%$ of draws in the games \citep[Sec~3.2]{Szczecinski20}.  
\end{itemize}

Finally, it is interesting to note that the \gls{fifa} rating algorithm uses score set $\tilde\mcY_\tr{FIFA}=\set{0,\frac{1}{2},\frac{3}{4},1}$, \citep{fifa_rating}, attributing $\tilde{y}=\frac{3}{4}$ to the user winning the game in a shootout. Since the  \gls{fifa} rating algorithm is not  based on a model,\footnote{Presumably, of course, because the FIFA rating does not explain what the origins of its rating are.} it cannot be directly related to the \gls{gelo} rating. In particular, it is because  the function $G(z)$ used in the \gls{fifa} rating algorithm is the same as in the Elo algorithm for the binary games and this, despite the fact, that the \gls{fifa} rating considers quaternary outcomes.

\section{Coefficients of the model}\label{Sec:Coeffs}

The generalization we propose comes  with increased complexity of the model which decreases the ``transparency'' as the operation of the rating algorithm depends on the coefficients \mbox{$\balpha=[\alpha_0,\ld, \alpha_J]$} and $\bdelta=[\delta_0,\ld,\delta_J]$ that must be found (estimated) from the data.

In that perspective, it may be argued that the popularity of the simple rating algorithms, such as the Elo algorithm, is mostly earned due to their transparency: since there are no coefficients to be found, the operation of the algorithm is predefined. However, even in this simplest case we run into trouble when we try to take the \acrfull{hfa} effect into account; this is preponderantly done by artificially increasing the rating of the home team by a constant factor 
\begin{align}\label{eta.theta.ij}
\theta_{t,i_t}\leftarrow\theta_{t,i_t}+\eta\sigma,
\end{align}
where the \gls{hfa} coefficient, $\eta$, is independent of the scale $\sigma$. 

While conceptually simple, this approach immediate raises a question of how the \gls{hfa} coefficient, $\eta$, should be found.\footnote{We note that the \gls{fifa} rating does not model the \gls{hfa} effect at all. This avoids the potentially controversial issue but the modeling is deficient.} This is fundamentally the same question we have to ask regarding the coefficients $\balpha$ and $\bdelta$ in our model.

Namely, i) should we do it in real-time, that is, treating the coefficients $\balpha$, $\bdelta$, and $\eta$ as unknowns which must be estimated in the same manner as the rating levels $\btheta_t$ are? or ii) should we do it off-line, estimating the coefficients from the historical data and then simply use these off-line estimates in the rating algorithm (such as the one we proposed in \secref{Sec:rating.filtering})?

Our answer to the above questions will depend on the purpose of rating. If it is to be used as a prediction tool as in \citet{Arntzen20}, then the real-time calculation is not only acceptable but likely preferable as we find the coefficients which fit the data we are analysing. 

On the other hand, the rating might be used as a tool for classification which has serious consequences such as relegation/promotion of the teams (that is, moving teams from one league to another, as done in the association football leagues). Then, the off-line calculation based on the historical data might be preferred because it increases the transparency of the algorithm: the coefficients are known before all the games so the rating of the teams participating in the game $t$ will depend only on its result $\tilde{y}_t$ and \emph{not} on past games of \emph{other} teams,  which would be the case if the coefficients were estimated from all the games before the game $t$.

Whether we choose an off-line or a real-time approach, we have to decide how to estimate the coefficients and we will consider two simple methods for that. We will formulate them in the context of the off-line estimation based on historical data, but both methods are applicable to the on-line estimation as well.

\subsection{Estimation from the game results}\label{Sec:fitting}

We can treat the coefficients as unknowns, akin to the rating levels, and fitting the coefficients $\balpha$, $\bdelta$, and $\eta$ to the games results may be done via the \gls{ml} method, that is
\begin{align}\label{ML.cdeta}
[\hat{\boldsymbol{\alpha}},\hat{\boldsymbol{\delta}},\hat{\eta}]=\argmax_{\boldsymbol{\alpha},\boldsymbol{\delta},\eta} \sum_{s\in\mcS_\tr{train}} \max_{\btheta^{(s)}} \sum_{t=1}^{T} L^{(s)}_{t}(\btheta^{(s)};\boldsymbol{\alpha},\boldsymbol{\delta},\eta),
\end{align}
where $L^{(s)}_{t}(\cd)$ is the log-likelihood we show in \eqref{likelihood}, $T$ is the number of games in each season, $s$, from the training season set, $\mcS_\tr{train}$, and where we explicitly show the dependence on the coefficients $\boldsymbol{\alpha}$, $\boldsymbol{\delta}$, $\eta$ we want estimate.

The optimization is carried out over the skills $\btheta^{(s)}$ of the teams participating in the season, $s$, and the skills are assumed constant throughout each season. However, $\btheta^{(s)}$ are not the same for different $s$; this is because i) we assume that the skills vary from season to season, and/or ii) the teams are not the same when we change $s$. The latter may happen due to the relegation and promotion rules (applicable particularly if we use the association football data as we do here). On the other hand, the coefficients $\boldsymbol{\alpha}$, $\boldsymbol{\delta}$, and  $\eta$ are shared among seasons, \ie we want to find the coefficients which provide the best fit in all  seasons in $\mcS_\tr{train}$. 

We note that, although we assumed that the parameters $\delta_h$ are monotonically growing with $h$, such a constraint is not imposed in the optimization; we rather count on the ``natural'' structure in the data to reveal this trend. This is also one of the reasons why the training set should be sufficiently large to capture such a behavior.

\subsection{Estimation from the frequency of the categories}\label{Sec:frequency}

The estimation method we presented in \secref{Sec:fitting} may raise some concerns. Not only the numerical optimization is required, which increases the implementation burden, but the solutions obtained cannot be easily related to the nature of the sport/competition.

To go around these difficulties, \citet{Szczecinski20} proposed a simple strategy which exploits the frequencies of the game results (observed in the past games/seasons).  In particular,  the \gls{hfa} coefficient, $\eta$, is derived from the frequencies of the home and away wins, denoted respectively by $f_{\mfH}$ and $f_{\mfA}$, see \citet[Sec.~3.2]{Szczecinski20}.

In such a  case, we are required to know, \emph{before} running the rating algorithm, what the frequencies of the home/away wins are. Indeed, it may be argued that these values are constant locally (within season) and globally (across seasons). This is, of course, only an approximation,\footnote{For example, ten \gls{epl} seasons starting from 2009 have values $f_{\mfH}\in (0.41,0.51) $ and $f_{\mfA}\in(0.24,0.34)$.}  but has an undeniable virtue of transparency and this is what seems to be cared about in practice.

We will thus apply here the reasoning of \citet{Szczecinski20}, generalizing it to the case of multi-level games using the frequencies of the events $\set{y_t=h}$ 
\begin{align}\label{f_h}
f_h = \frac{1}{T|\mcS_\tr{train}|}\sum_{s\in\mcS_\tr{train}}\sum_{t=1}^{T}\IND{y^{(s)}_t=h},
\end{align}
where $\IND{\cd}$ is the indicator function and, following the notation from \secref{Sec:fitting}, $y^{(s)}_t$ denotes the outcome of the game in the season indexed by $s$. 

First, we rewrite \eqref{Pr.y.t.normalized} to include the \gls{hfa}
\begin{align}
\PR{Y_t=h|z_t} = P_h( z_t + \eta \sigma  )=
\frac{ 10^{\alpha_h+\delta_h\big(\frac{z_t}{\sigma}+\eta\big)}}
{\sum_{l=0}^J 10^{\alpha_l+ \delta_l\big(\frac{z_t}{\sigma}+\eta\big)}}, \quad h=0,\ld, J.
\end{align}


Next, we assume that teams strengths, $\theta_{t,i}, i=1,\ld,M$ do not change during a long period of time (\eg  the entire season with $T$ games), that is $\theta_{t,i}=\theta_i$, and we calculate the probability of observing the particular outcome $\set{Y_t=h}$ in that period
\begin{align}
\label{Pr.Y.h.i.j}
\PR{Y_t=h}&=\sum_{t=1}^T\PR{Y_t=h|i_t, j_t}\PR{i_t,j_t}\\
\label{Pr.Y.h}
&=\frac{1}{M(M-1)}\sum_{\substack{i,j=1\\i\neq j}}^M P_h( \theta_i-\theta_j +\eta\sigma) \\
\label{Pr.Y.h.2}
&\approx \frac{1}{M(M-1)}\sum_{\substack{i,j=1\\i\neq j}}^M \big[ P_h( \eta \sigma  ) + P'_h(\eta\sigma)(\theta_i-\theta_j)\big]\\
\label{Pr.Y.h.3}
&=P_h( \eta \sigma  ),
\end{align}
where  Eq.~\eqref{Pr.Y.h.i.j} assumes that the assignment of the teams $i_t$ and $j_t$ to the $t$-th game is random,  Eq.~\eqref{Pr.Y.h} assumes that the assignments are equiprobable,\footnote{For example, in the \gls{epl} each team plays every other team exactly twice, once at home and once away. In this case the summation over $T$ means the summation over the entire season. In other sports, the encounters are more frequent.}  and  Eq.~\eqref{Pr.Y.h.2} linearizes the probability function around $\theta_i-\theta_j=0$ which is valid if we assume that the differences $\theta_i-\theta_j$ are small comparing to $\sigma$. 

Finally, replacing the probability of the event defined in \eqref{Pr.Y.h.3} by its estimator from \eqref{f_h} we obtain the set of $J+1$ equations
\begin{align}
\xi 10^{\alpha_h+\delta_h\eta} & = f_h, \quad h=0,\ld,J,
\end{align} 
where $\xi=\left[\sum_{l=0}^J 10^{\alpha_l+\delta_l\eta}\right]^{-1}$ is the normalization factor.

From the symmetry conditions $\alpha_h=\alpha_{J-h}$, and $\delta_h+\delta_{J-h}=0$, see  \eqref{alpha.delta}, the equations should be solved in pairs
\begin{align}
\label{first.line}
\xi 10^{\alpha_0+\delta_0\eta} & = f_0,\quad &&&\xi 10^{\alpha_J+\delta_J\eta} & = f_J,\\
\xi 10^{\alpha_1+\delta_1\eta} & = f_1,\quad &&&\xi10^{\alpha_{J-1}+\delta_{J-1}\eta} & = f_{J-1},\\
&\vd&&&&\vd\\
\label{last.line}
&&\xi \cd 10^{\alpha_{J/2}}   = f_{J/2},
\end{align}
where only one equation appears in \eqref{last.line} because we assumed $J$ to be even (in which case we also know that $\delta_{J/2}=0$). If $J$ is odd, we have to replace \eqref{last.line} with
\begin{align}
\label{last.line.off}
\xi 10^{\alpha_{(J-1)/2}+\delta_{(J-1)/2} \cd\eta} & = f_{(J-1)/2},\quad &\xi 10^{\alpha_{(J+1)/2}+\delta_{(J+1)/2}\cd \eta} & = f_{(J+1)/2}.
\end{align}

Since we assumed $\alpha_0=\alpha_J=0$ and $\delta_J= -\delta_{0}=1$, the parameters $\xi$ and $\eta$ are calculated as
\begin{align}
\label{xi.total}
\xi &=\sqrt{f_0 f_J},\\
\label{eta.total}
\eta & = \frac{1}{2} \log_{10}\frac{f_{J}}{f_0}.
\end{align}

Then, to solve  Eqs.~\eqref{first.line}-\eqref{last.line}, we multiply equations in the same line which yields
\begin{align}\label{xi.c}
\alpha_h 
& = \frac{1}{2}\log_{10}(f_h f_{J-h} ) -\log_{10}\xi, \quad h=1,\ld, J-1,
\end{align}
by dividing them we get 
\begin{align}\label{delta.J}
 \delta_{h}  = \frac{1}{2\eta} \log_{10}\frac{f_{h}}{f_{J-h}}, \quad h=1,\ld, J-1,
\end{align}
and we note that, when $J$ is even,  we automatically obtain $\delta_{J/2}=0$ (which is a consequence of the relationship given by Eq.~\eqref{alpha.delta}) and $\alpha_{J/2}  = \log_{10} f_{J/2} -\log_{10}\xi$, which solves Eq.~\eqref{last.line}.

Some cautionary statements are in order:
\begin{itemize}
\item Since $f_h$ are estimated from the observations, a care should be taken to guarantee that a sufficient number of results in a particular category, $h$, is available.  In particular, the above equations make sense only if $f_h>0, h=0, \ld, J$. Gathering the statistics from multiple seasons is useful with that regard.
\item The relationship \eqref{delta.J} is valid only for $\eta\neq 0$; if $f_0=f_J$ and thus $\eta=0$ we cannot estimate $\delta_h$. In more general terms,  the dependence of all the parameters $\delta_h$ on the quality of estimation of $\eta$ via the relationship \eqref{eta.total} is a delicate point and deserves a further investigation. 
\item Similarly to the optimization approach, there are no constraints imposed a monotonic behavior of $\delta_h$ and the results we obtained were ``naturally'' appearing from the collected data but clearly, the issue of finding $\bdelta$ should be analyzed in more depth.
 \end{itemize}
 
As a sanity check, we can compare these equations to those presented in \citet{Szczecinski20} (\ie for a particular case of $J=2$), where  only the \gls{hfa} parameter, $\eta$, and the parameter related to the draws, $\kappa=10^{\alpha_1}$ need to be estimated; they are obtained from \eqref{eta.total}  and  \eqref{delta.J}
\begin{align}
\label{eta.J=2}
\eta&= \frac{1}{2} \log_{10}\frac{f_{\mfH}}{f_\mfA},\\
\label{c1.J=2}
\alpha_1  &= \log_{10}\frac{f_\mfD}{\sqrt{f_\mfH f_\mfA}},
\end{align}
where, for compatibility with \citet{Szczecinski20}, we used $f_0=f_\mfA$, $f_1=f_\mfD$, and $f_2=f_\mfH$. 

Indeed, these equations correspond to \citet[Eqs.~(44)-(45)]{Szczecinski20}. The only difference resides in the factor $\frac{1}{2}$ in \eqref{eta.J=2} which is due to doubling of the scale $\sigma$ used in \citet[Eqs.~(24)-(26)]{Szczecinski20}.


\section{Numerical Examples}\label{Sec:Examples}
We use here the data from two  team sports: 
\begin{itemize}
\item The \gls{epl} association football, where, in each season $M=20$ teams face each other in one home and one away games; there are thus $T=M(M-1)=380$ games in total;
\item The \gls{nfl} American football, where, in each season $M=32$ teams play 16 games each according to the predefined schedule; there are thus $T=16M/2=256$ games in total.
\end{itemize}

These sports are quite different regarding the possible definition of the \gls{mov}: in the association football, the goal difference is a relatively small value and there are several works which postulated the use of predefined distributions to model this difference directly, see \secref{Sec:MOV.literature}. On the other hand, in the case of the \gls{nfl} games, the points scored correspond to a variety of events (such as, for example, a \emph{touchdown} worth six points or a \emph{try} after the touchdown which is worth one point). In this case, modeling the points directly would be challenging and the discretization is particularly useful because it lumps together all  the points.

Data is taken from ten consecutive seasons starting 2009/10 which were retrieved from \citet{football-data} (\gls{epl}) and from \citet{NFL-reference} (\gls{nfl}).  
The first five ``training'' seasons 2009/10 -- 2013/14 form $\mcS_\tr{train}$ while the last five seasons,  2014/15 -- 2018/19, are gathered in the testing set $\mcS_\tr{test}$ and will be used to evaluate the algorithms; the same approach was adopted in \citet{Arntzen20}. The training is thus based on the results of $5\cd 380 = 1900$ games for the \gls{epl} and of $5\cd 256=1280$ games for the \gls{nfl}.

We  study the model \eqref{res.sA}-\eqref{res.sJ}, for three different values of thresholds $\Delta$ and also consider its generalization to $J+1=7$ discretization categories defined as  $\set{d_t<-\Delta''}$, $\set{-\Delta''\le d_t<-\Delta'}$, $\set{-\Delta'\le d_t<0}$, $\set{d_t=0}$, $ \set{0< d_t\le\Delta'}$, $\set{\Delta'< d_t\le-\Delta''}$, and $ \set{\Delta''< d_t}$, indexed, respectively by $y_t=0,\ld, 6$.

Although we ensured that the values $f_h$ are far from zero, see \tabref{Table.f.h}, no formal criterion was used to find the thresholds as our goal was rather to illustrate the flexibility of the proposed algorithm. For the \gls{epl} we use $\Delta\in\set{1,2,3}$ (when $J+1=5$) as well as $\Delta'=1$ and $\Delta''=2$ ($J+1=7$).  For the \gls{nfl} the difference in the scores are larger than in association football and we use $\Delta\in\set{5,10,15}$ ($J+1-5$) and $\Delta'=5$ and $\Delta''=10$ ($J+1=7$).

In the first step, we need to find the coefficients of the models. If we opt for the frequency-based estimation of \secref{Sec:frequency}, we first calculate the frequencies $f_h$,  \eqref{f_h}, which  are  shown in \tabref{Table.f.h}; note that the frequency of the draw $f_\mfD=f_{J/2}$ is always the same. In the same manner,  the frequencies of away- and home wins are calculated respectively as $f_\mfA=\sum_{h=0}^{J/2-1}f_h$ and $f_\mfH=\sum_{h=J/2+1}^J f_h$.

\begin{table}[tb]
\centering
\scalebox{0.8}{
\begin{tabular}{c|c|| c|c|c|c || c|c}

\multicolumn{2}{c||}{}	&   		 \multicolumn{4}{c||}{$J=4$}	&   \multicolumn{2}{c}{$J=6$}\\
		\cline{3-6}
\multicolumn{2}{c||}{$J=2$}  &   		& $\Delta=1$ 		&  $\Delta=2$ 	&  $\Delta=3$ 	&   \multicolumn{2}{c}{$\Delta'=1, \Delta''=2$}\\
			\hline
$f_0$ 		& $0.277$  & $f_0$	&  $0.127$ 		&  $0.051$ 		&   $0.018$		& $f_0$	& $0.051$\\
$f_1$ 		& $0.256$  & $f_1$	&  $0.150$ 		&  $0.226$ 		&   $0.258$		& $f_1$	& $0.076$\\
$f_2$ 		& $0.467$  & $f_2$	&  $0.256$ 		&  $0.256$ 		&   $0.256$		& $f_2$	& $0.150$\\
		&	& $f_3$	&  $0.219$ 		&  $0.353$ 		&   $0.420$		& $f_3$	& $0.256$\\
		&	& $f_4$	&  $0.248$ 		&  $0.114$ 		&   $0.047$		& $f_4$	& $0.219$\\
		&	  & 		&  			&  			&   			& $f_5$	& $0.134$\\
		&	  & 		&  			&  			&   			& $f_6$	& $0.114$
\end{tabular}
}

a) \gls{epl}
\vskip 0.2cm

\scalebox{0.8}{
\begin{tabular}{c|c|| c|c|c|c || c|c}

\multicolumn{2}{c||}{}	&   		 \multicolumn{4}{c||}{$J=4$}	&   \multicolumn{2}{c}{$J=6$}\\
		\cline{3-6}
\multicolumn{2}{c||}{$J=2$}  &   	& $\Delta=5$ 	&  $\Delta=10$ 	&  $\Delta=15$ 	&   \multicolumn{2}{c}{$\Delta'=5, \Delta''=10$}\\
			\hline
$f_0$ 		& $0.426$  & $f_0$	&  $0.280$ 		&  $0.166$ 		&   $0.116$		& $f_0$	& $0.166$\\
$f_1$ 		& $0.001$  & $f_1$	&  $0.146$ 		&  $0.259$ 		&   $0.309$		& $f_1$	& $0.113$\\
$f_2$ 		& $0.573$  & $f_2$	&  $0.001$ 		&  $0.001$ 		&   $0.001$		& $f_2$	& $0.146$\\
		&		& $f_3$	&  $0.169$ 		&  $0.311$ 		&   $0.381$		& $f_3$	& $0.001$\\
		&		& $f_4$	&  $0.404$ 		&  $0.262$ 		&   $0.191$		& $f_4$	& $0.169$\\
		&	  & 		&  			&  			&   			& $f_5$	& $0.142$\\
		&	  & 		&  			&  			&   			& $f_6$	& $0.262$
\end{tabular}
}

b)  \gls{nfl}
\caption{The frequencies, $f_h, h=0,\ld,J$,  of the events $\set{y_t=h}$ calculated over five consecutive  seasons starting from 2009/10 for  a) \gls{epl}, and b) \gls{nfl}.}\label{Table.f.h}
\end{table}

Next, by applying the formulas \eqref{xi.total}-\eqref{delta.J} to the frequencies of the categories, $f_h$, we  estimate the coefficients of the model and show them in  \tabref{Table.c.delta.freq}, where we also show the coefficients estimated by solving the optimization problem \eqref{ML.cdeta}.  Once the coefficients $\balpha$, $\bdelta$, and $\eta$ are estimated, $\tilde{K}$  (which controls the amplitude of the updates, see \eqref{algo.i}-\eqref{algo.j}) is found so as to maximize the  average logarithmic scores defined in \eqref{LS}, where instead of $\mcS_\tr{test}$, we use the training seasons $\mcS_\tr{train}$; this calibration of the algorithm is thus done on the same historical data that was used to estimate the coefficients $\balpha$, $\bdelta$, and $\eta$. The coefficient $\tilde{K}$ is also show in  \tabref{Table.c.delta.freq}.

 The Elo-Davidson and the \gls{gelo} algorithms are then applied to the data from testing seasons in $\mcS_\tr{test}$, and since their coefficients are defined by \tabref{Table.c.delta.freq},  the operation of the rating algorithms is entirely transparent. 

\begin{table}
\centering
\scalebox{0.8}{
\begin{tabular}{c|c|c|| c|c|c|c|c|c|c || c|c|c}

\multicolumn{3}{c||}{}	&   		 \multicolumn{7}{c||}{$J=4$}	&   \multicolumn{3}{c}{$J=6$}\\
		\cline{4-10}
\multicolumn{3}{c||}{$J=2$}  &   			& \multicolumn{2}{c|}{$\Delta=1$} 	&  \multicolumn{2}{c|}{$\Delta=2$} 	&  \multicolumn{2}{c||}{$\Delta=3$}	&   \multicolumn{3}{c}{$\Delta'=1, \Delta''=2$}\\
			\hline
		 	&  Freq.  	& Opt. & 		 		&   Freq.  	& Opt.	&   Freq.  	& Opt.	&    Freq.  	& Opt.	& 				&  Freq.  & Opt.\\
			\hline
$\alpha_1$ 	& $-0.15$  	& $-0.06$ 	& $\alpha_1$ 	&  $0.01$ 	& $0.18$	&  $0.57$ 	& $0.85$	&   $1.05$	&$1.47$	& $\alpha_1$		& $0.12$&$0.34$\\
$\eta$ 		& $0.11$  	& $0.15$	& $\alpha_2$	&  $0.16$ 	& $0.35$	&  $0.53$ 	& $0.86$	&   $0.94$	& $1.42$	& $\alpha_2$		& 	$0.38$& $0.68$\\
$\tilde K$ 		& $0.06$  	& $0.07$	& $\tilde\delta_1$&  $0.22$ 	& $0.31$	&  $0.22$ 	& $0.30$	&   $0.24$	&$0.31$	& $\alpha_3$		& $0.53$&$0.86$\\
			&		&		& $\eta$		&  $0.15$ 	& $0.21$ 	&  $0.17$ 	& $0.27$	&   $0.21$	&$0.34$	& $\tilde\delta_1$	& $0.15$&$0.20$\\
			&		&		& $\tilde K$	&  $0.10$ 	& $0.14$	&  $0.14$ 	& $0.24$	&   $0.20$	&$0.35$	& $\tilde\delta_2$	& $0.27$&$0.35$\\
			&	  	&		& 			&  		&		&  		&		&   		&		& $\eta$			& $0.17$&$0.27$\\
			&	  	&		& 			&  		&		&  		&		&   		&		& $\tilde K$		& $0.14$&$0.24$
\end{tabular}
}

a) \gls{epl}
\vskip 0.2cm

\scalebox{0.8}{
\begin{tabular}{c|c|c|| c|c|c|c|c|c|c || c|c|c}

\multicolumn{3}{c||}{}	&   		 \multicolumn{7}{c||}{$J=4$}	&   \multicolumn{3}{c}{$J=6$}\\
		\cline{4-10}
\multicolumn{3}{c||}{$J=2$}  &   			& \multicolumn{2}{c|}{$\Delta=5$} 	&  \multicolumn{2}{c|}{$\Delta=10$} 	&  \multicolumn{2}{c||}{$\Delta=15$}	&   \multicolumn{3}{c}{$\Delta'=5, \Delta''=10$}\\
			\hline
		 	&  Freq.  	& Opt. & 		 		&   Freq.  	& Opt.	&   Freq.  	& Opt.	&    Freq.  	& Opt.	& 				&  Freq.  & Opt.\\
			\hline
$\alpha_1$ 	& $-2.50$  	& $-2.41$ 	& $\alpha_1$ 	&  $-0.33$ 	& $-0.17$		&  $0.13$ 	& $0.13$		&   $0.36$	&$0.64$		& $\alpha_1$		& $-0.21$	&$-0.04$\\
$\eta$ 		& $0.06$  	& $0.09$	& $\alpha_2$	&  $-2.33$ 	& $-2.17$		&  $-2.13$ 	& $-2.13$		&   $-1.98$	& $-1.65$		& $\alpha_2$		& $-0.12$	&$0.11$\\
$\tilde K$ 		& $0.07$  	& $0.07$	& $\tilde\delta_1$&  $0.30$ 	& $0.40$		&  $0.30$ 	& $0.30$		&   $0.29$	&$0.34$		& $\alpha_3$		& $-2.12$	&$-1.88$\\
			&		&		& $\eta$		&  $0.08$ 	& $0.12$ 		&  $0.10$ 	& $0.10$		&   $0.11$	&$0.18$		& $\tilde\delta_1$	& $0.25$	&$0.24$\\
			&		&		& $\tilde K$	&  $0.10$ 	& $0.13$		&  $0.15$ 	& $0.20$		&   $0.19$	&$0.27$		& $\tilde\delta_2$	& $0.34$	&$0.42$\\
			&	  	&		& 			&  		&			&  		&			&   		&			& $\eta$			& $0.10$	&$0.16$\\
			&	  	&		& 			&  		&			&  		&			&   		&			& $\tilde K$		& $0.15$	&$0.20$
\end{tabular}
}

b) \gls{nfl}

\caption{Coefficients of the Elo-Davidson ($J+1=3$ discretization levels) and the \gls{gelo} ($J+1=5$ and $J+1=7$ discretization levels) algorithms, obtained from the training seasons in a) \gls{epl} and b) \gls{nfl}. ``Freq.'' means that the coefficients were obtained by applying  Eqs.~\eqref{xi.total}-\eqref{delta.J} to the frequencies of the categories, $f_h, h=0,\ld,J$ shown in \tabref{Table.f.h}, while ``Opt.'' means that they were obtained solving the optimization problem in \eqref{ML.cdeta}. The coefficients not shown in the table are obtained by definition and/or symmetry: $\alpha_0=\alpha_J=0$, $\tilde\delta_0=0$, $\tilde\delta_J=1$, and $\tilde\delta_{J/2}=\frac{1}{2}$, $\alpha_{J-h}=\alpha_h$,  $\tilde\delta_{J-h}=1-\tilde\delta_h$. The normalized step, $\tilde{K}$, is found empirically by minimizing the logarithmic score, \eqref{LS}, in the \emph{training} seasons, that is using $\mcS_\tr{train}$ rather than $\mcS_\tr{test}$.}\label{Table.c.delta.freq}
\end{table}

To compare the results obtained with different models (including different coefficients estimation strategies),  we will evaluate the predictive strength of the underlying models.  This is possible if we limit out comparison to the ternary results and, since in all the algorithms we defined the draw in the same manner, the event of the home win ($\mfH$) is obtained by merging all the events for which $d_t>0$; the same is done for the away win ($\mfA$) which yields the prediction of the home win, away win, and draw events
\begin{align}
P_{\mfA}(z_t) &= \sum_{h=0}^{J/2-1}P_{h}(z_t+\eta\sigma), \\
P_{\mfD}(z_t) &=P_{J/2}(z_t+\eta\sigma),\\
P_{\mfH}(z_t) &= \sum_{h=J/2+1}^J P_{h}(z_t+\eta\sigma),
\end{align}
which are then used to calculate the following performance metrics season-indexed with $s$: 
\begin{itemize}
\item The logarithmic score 
\begin{align}\label{metric:logloss}
\tr{LS}_t^{(s)}= -\log P_\mfA(z_t)  \IND{\mfA}  - \log P_\mfD(z_t)  \IND{\mfD} -\log P_\mfH(z_t)  \IND{\mfH},
\end{align}
which is often applied to evaluate the quality of prediction \citep{Gelman14} but ignores the ordinal nature of the outcomes;
\item The \gls{rps}, which takes into account the ordinality of the game results \citep[Sec.~3]{Constantinou12}, \citep[Sec.~3.3]{Lasek20}; it may be written as\begin{align}\label{metric:rps}
\tr{RPS}_t^{(s)}= \frac{1}{2}\big[(P_\mfA(z_t) - \IND{\mfA})^2  +(P_\mfA(z_t)+P_\mfD(z_t) -\IND{\mfA\vee \mfD})^2\big];
\end{align}
\item The accuracy 
\begin{align}\label{metric:ac}
\tr{AC}_t^{(s)}= \IND{\argmax_{x\in\set{\mfA,\mfD,\mfH}}P_x(z_t)},
\end{align}
which says if the event with the maximum predicted probability was actually observed.
\end{itemize}

Finally, we average the metrics over the period of interest in each of the testing seasons
\begin{align}\label{LS}
\ov{\tr{LS}}&=\frac{1}{(T-\tau)|\mcS_\tr{test}|}\sum_{s\in\mcS_\tr{test}}\sum_{t=\tau+1}^T \tr{LS}_t^{(s)},\\
\label{RPS}
\ov{\tr{RPS}}&=\frac{1}{(T-\tau)|\mcS_\tr{test}|}\sum_{s\in\mcS_\tr{test}}\sum_{t=\tau+1}^T \tr{RPS}_t^{(s)},\\
\label{AC}
\ov{\tr{AC}}&=\frac{1}{(T-\tau)|\mcS_\tr{test}|}\sum_{s\in\mcS_\tr{test}}\sum_{t=\tau+1}^T \tr{AC}_t^{(s)},
\end{align}

Since the rating is based on the \gls{sg} algorithm, which is known to converge slowly to the solution, we  assume that the convergence occurs after the game  $\tau=\frac{1}{2}T$, that is, the averages \eqref{LS}-\eqref{AC} are taken over the second half of the seasons.

\begin{table}
\centering

\scalebox{0.8}{
\begin{tabular}{c||c|c|| c|c|c|c|c|c || c|c||c}
&\multicolumn{2}{c||}{}	&    \multicolumn{6}{c||}{$J=4$}	&   \multicolumn{2}{c||}{$J=6$}\\
		\cline{4-9}
&\multicolumn{2}{c||}{$J=2$}  &   	\multicolumn{2}{c|}{$\Delta=1$} 	&  \multicolumn{2}{c|}{$\Delta=2$} &  \multicolumn{2}{c||}{$\Delta=3$}	&   \multicolumn{2}{c||}{$\Delta'=1, \Delta''=2$}& $f_{\mfA/\mfD/\mfH}$\\
			&  Freq.  		& Opt. 	& 	Freq.  		& Opt.		&   Freq.  		& Opt.		&    Freq.  		& Opt.	& 	Freq.  & Opt.\\
			\hline
$\ov{\tr{LS}}$ 	& $0.9740$  	& $0.9785$ 	&   $0.9696$ 	& $0.9716$		&  $0.9690$ 	& $0.9710$		&   $0.9703$	&$0.9724$	&  $\mb{0.9679}$	& $0.9695$ & $1.0540$\\
$\ov{\tr{RPS}}$ 	& $0.2006$  	& $0.2010$		&  $0.1993$ 	& $0.1991$		&  $0.1990$ 	& $0.1988$		&   $0.1995$	& $0.1993$	&  $0.1987$	& $\mb{0.1984}$ & $0.2281$\\
$\ov{\tr{AC}}$ 	& $0.5442$  	& $0.5442$		&   $0.5432$ 	& $0.5442$		&  $0.5421$ 	& $\mb{0.5453}$		&   $0.5411$	&$0.5400$	&  $0.5389$ &	$0.5389$ & $0.4758$
\end{tabular}
}

a) \gls{epl}

\vskip 0.2cm

\scalebox{0.8}{
\begin{tabular}{c||c|c|| c|c|c|c|c|c || c|c||c}
&\multicolumn{2}{c||}{}	&    \multicolumn{6}{c||}{$J=4$}	&   \multicolumn{2}{c||}{$J=6$}\\
		\cline{4-9}
&\multicolumn{2}{c||}{$J=2$}  &   	\multicolumn{2}{c|}{$\Delta=4$} 	&  \multicolumn{2}{c|}{$\Delta=10$} &  \multicolumn{2}{c||}{$\Delta=15$}	&   \multicolumn{2}{c||}{$\Delta'=5, \Delta''=10$} & $f_{\mfA/\mfD/\mfH}$\\
			&  Freq.  	& Opt. 	& 	Freq.  	& Opt.	&   Freq.  	& Opt.	&    Freq.  			& Opt.	& Freq.  	& Opt. \\
			\hline
$\ov{\tr{LS}}$ 	& $0.6304$  & $0.6335$ &   $0.6264$ & $0.6287$	&  $0.6224$ & $0.6238$&   $\mb{0.6223}$	&$0.6231$	&  $0.6224$	&$0.6242$ & $0.6881$\\
$\ov{\tr{RPS}}$ 	& $0.2200$  & $0.2214$&  $0.2182$ & $0.2193$	&  $0.2166$ & $0.2173$&   $\mb{0.2162}$	& $0.2164$	&  $0.2166$	& $0.2174$ & $0.2467$\\
$\ov{\tr{AC}}$ 	& $0.6375$  & $0.6250$&   $0.6469$ & $0.6391$	&  $0.6531$ & $0.6469$&   $0.6516$		&$0.6453$	&  $\mb{0.6656}$	&$0.6531$ & $0.5594$
\end{tabular}
}

b) \gls{nfl}

\caption{The metrics  \eqref{LS}-\eqref{AC} obtained in a) \gls{epl} and b) \gls{nfl} using the \gls{gelo} algorithms specified in \tabref{Table.c.delta.freq}. The results in the column $f_{\mfA/\mfD/\mfH}$ are obtained assuming non-algorithmic prediction for all the games based on the frequencies of the away-win ($f_\mfA$), draw ($f_\mfD$), and home-win ($f_\mfH$) events (obtained from $\mcS_\tr{train})$, that is assuming $P_x(z)=f_x, x\in\set{\mfA, \mfD, \mfH}$. The best numerical values are emphasized with bold font.}\label{Tab:metrics}
\end{table}

\begin{figure}
\centering
\psfrag{title}{}
\psfrag{Elo-XXX-XXX-XXX}{\footnotesize Elo-Davidson}
\psfrag{BET365}{\footnotesize Bet365}
\psfrag{1-}[lb]{\footnotesize $\Delta=1$}
\psfrag{2-}[lb]{\footnotesize  $\Delta=2$}
\psfrag{3-}[lb]{\footnotesize $\Delta=3$}
\psfrag{1-2-}[lb]{\footnotesize $\Delta'=1$, $\Delta''=2$}

\psfrag{xlabel}[c][c]{\footnotesize $\tilde{K}$}
\psfrag{ylabel}[b][c]{\footnotesize $\ov{\tr{LS}}$}
\includegraphics[width=0.85\linewidth]{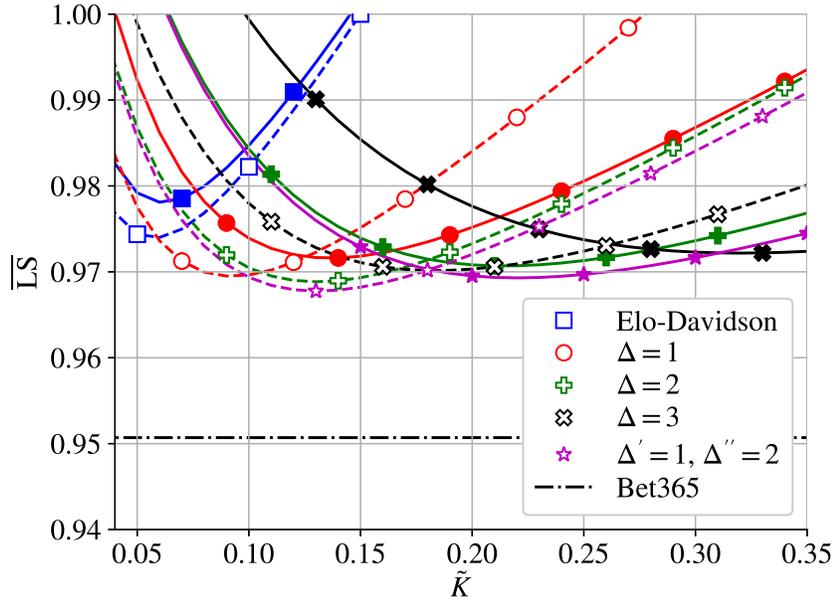}

\caption{Logarithmic score defined in \eqref{LS} as a function of $\tilde{K}$ for the Elo-Davidson and the \gls{gelo} algorithms (the later, for different discretization thresholds, $\Delta$). The coefficients $\balpha$, $\bdelta$, and $\eta$ used in the algorithms are obtained via frequency-based formulas from \secref{Sec:frequency} (dashed lines, hollow markers) or by optimization method from \secref{Sec:fitting} (solid lines, filled markers). The logarithm score ``Bet365'' is based on the probabilities inferred from the betting odds offered by the site Bet365 (by inverting and normalizing the decimal odds).}
\label{Fig:LogScore}
\end{figure}

As an indication that the coefficients of the model obtained from the  training data can be reliably used on the testing set, we show in \figref{Fig:LogScore} the logarithmic score and we can appreciate that $\tilde{K}$ found from $\mcS_\tr{train}$ (shown in \tabref{Table.c.delta.freq}) also minimizes  $\ov{\tr{LS}}$ when the algorithms are deployed on $\mcS_\tr{test}$ (as can be appreciated in \figref{Fig:LogScore}). 

\tabref{Tab:metrics} gathers all the metrics calculated for different models and we can appreciate that, 
\begin{itemize}
	\item By increasing the complexity of the model, \ie  by using the \gls{ac}-based \gls{mov} modeling, we are able to improve the prediction results comparing to the Elo-Davidson algorithm.  By no means it should be considered an obvious outcome: while the model with a larger number of coefficients ($\balpha$, $\bdelta$) must improve the fit to the training data, there is no guarantee that the fit to the testing set will improve as well.  The improved prediction capability is thus the evidence of a the model suitably fitting the data in the sports we considered here. 
	\item Although the differences in the logarithmic scores, $\ov{\tr{LS}}$, seem rather small, they translate into the improved accuracy, $\ov{\tr{AC}}$. The improvement over the non-algorithmic prediction (based on the prior frequencies $f_\mfA, f_\mfD, f_\mfH$) is clear already for the Elo-Davidson algorithm (\ie for $J=2$), where we gain approximately $7\%$ for both the \gls{epl} and the \gls{nfl}. On the other hand, further increase in the complexity of the model ($J=4$ or $J=6$) brings very small change in the prediction accuracy for the \gls{epl}; in the \gls{nfl}, however, additional $3\%$ are gained.\footnote{These small changes are important. As a reference, we note that beating the bookmakers' prediction by couple of percents may be sufficient to ensure the monetary gains.} The difference in the prediction accuracy between both sports is due  to the fact that the \gls{nfl} games are practically binary  (with extremely low probability of draws) and, ignoring the draws we can notably decrease the prediction errors.
	\item 
Comparing the frequency-based and the optimization-based coefficient estimation methods, it seems that the former has a (slightly) better generalization properties, \ie fits better the independent data. We do not have a formal explanation for these results but we conjecture that the optimization leads to overfitting while the frequency-based estimation, being decoupled (by averaging defined in \eqref{f_h}) from the game results, regularizes the estimates; however,  this issue  deserves further investigation. 
\end{itemize}	

We also note that the behaviour in \figref{Fig:LogScore} is not necessarily the same when algorithms are evaluated over individual testing seasons. However, since we opted for the evaluation of the generalization capability of the algorithm, taking the average result as we do in \eqref{LS}-\eqref{AC}, is the appropriate methodology to assess the gains. If, on the other hand, the prediction is the main goal, further improvements may be sought by adjusting the coefficients of the model using data from the season being evaluated; see the discussion in \secref{Sec:Coeffs}.

\section{Conclusions}\label{Sec:Conclusions}

In this work, we proposed a new rating algorithm based on a probabilistic model of the relationship between the skills of the teams and the discretized outcomes of the games. Using higher-than-ternary discretization we thus addressed the problem of using the \acrfull{mov} in the rating algorithms. 

The \acrfull{ac} model we propose to use is shown to be a natural generalization of the model underlying the Elo algorithm. Using the \acrfull{sg} to solve the \acrfull{ml} estimation we derive the \acrfull{gelo} algorithm. Both, the Elo and the \gls{gelo} algorithms, have essentially the same update equations but, the latter takes the multi-level games results into account, by redefining the score and the expected score. 

We apply the \gls{gelo} algorithm to the association football and American football data of the \gls{epl} and the \gls{nfl} leagues, and observe the impact of discretization strategies on the gain offered by the \gls{gelo} algorithm over the Elo-Davidson algorithm based on the conventional ternary-outcome discretization. The improvements, although small, are notable indicating that the \gls{ac} model is a suitable choice to rate the teams in the analyzed sports.


Furthermore, we show that the model can be easily adjusted to fit the statistics of the targeted sport. More specifically,  the coefficients of the model  can be found in closed-form from the frequencies with which the discretized \gls{mov} results are observed. The advantage of using this approach is discussed comparing to the alternative, optimization strategy. While the proposed method yields the algorithms which easily and transparently take into account the sport-specificity, it has some pitfalls we also identified. In particular, more research is necessary to find simple and  robust methods to defining the discretization thresholds and to estimate the coefficients of the model.

Finally, the \gls{mov}-based  rating algorithm we proposed may be applied and tested in other popular sports where the ranking plays an important role. For example, it  might be combined with the in-play points in tennis, see \cite{Kovalchik16} for a thorough overview, or it may be used to evaluate the contribution of the individual players in the team sports, \cite{Gramacy13}, \cite{Hvattum21},  where the \gls{mov} is also used, \cite{Wolf20}.


\end{document}